\documentclass[structuredabstract]{aa}

\usepackage{graphicx} 
\usepackage{subfig} 
\usepackage[round]{natbib}
\usepackage{color}
\usepackage{comment}
\usepackage{amsmath}
\usepackage{fixltx2e}
\usepackage{txfonts}

\newcommand{\Omegab}{\Omega_{\mathrm{b}}}

\newcommand{\RCR}{R_{\mathrm{CR}}}
\newcommand{\ROLR}{R_{\mathrm{OLR}}}
\newcommand{\vOLR}{v_{\mathrm{OLR}}}

\newcommand{\Kpc}{~\mathrm{kpc}}
\newcommand{\pc}{~\mathrm{pc}} 
\newcommand{\Gyr}{~\mathrm{Gyr}}

\newcommand{\kmsec}{~\mathrm{km}~\mathrm{s}^{-1}}
\newcommand{\kmseckpc}{~\mathrm{km}~\mathrm{s}^{-1}~\mathrm{kpc}^{-1}}

\newcommand{\Msun}{M_{\odot}}

\newcommand{\Mbar}{M_{\mathrm{bar}}} 

\newcommand{\de}{\mathrm{d}}

\newcommand{\Tbar}{T_\mathrm{bar}}

\newcommand{\Rg}{R_\mathrm{g}}

\newcommand{\RNum}[1]{\uppercase\expandafter{\romannumeral #1\relax}}

\newcommand{\Nthin}{N_\mathrm{thin}}
\newcommand{\Nthick}{N_\mathrm{thick}}
\newcommand{\MK}{M_{K}}
\newcommand{\degree}{^{\circ}}
\newcommand{\KBB}{K_\mathrm{BB}}
\newcommand{\JBB}{J_\mathrm{BB}}
\newcommand{\mg}{\ \mathrm{mag}}

\newcommand{\epm}{e_\mathrm{p.m.}}
\newcommand{\elos}{e_\mathrm{los}}
\newcommand{\vlos}{v_\mathrm{los}}
\newcommand{\avvR}{\overline{v}_R}
\newcommand{\avvphi}{\overline{v}_\phi}
\newcommand{\avvz}{\overline{v}_z}
\newcommand{\RAVE}{\emph{RAVE} }

\newcommand{\Nb}{N_\mathrm{b}}

\begin{document}
\title{The Galactic bar and the large scale velocity gradients in the
  Galactic disk}
\titlerunning{The Galactic bar and the large scale velocity gradients}

\author{G. Monari \inst{1} \and A. Helmi \inst{1} \and T. Antoja
  \inst{1}\fnmsep\inst{2} \and M. Steinmetz \inst{3} }

\institute{Kapteyn Astronomical Institute, Rijksuniversiteit
  Groningen, P.O. Box 800, 9700 AV Groningen, The Netherlands \and
  ESA, European Space Research and Technology Center (ESTEC),
  Keplerlaan 1, 2201 AZ Noordwijk, The Netherlands \and
  Leibniz-Institut f\"{u}r Astrophysik Potsdam (AIP), An der
  Sternwarte 16, 14482 Potsdam, Germany \\
  \email{monari@astro.rug.nl}}

\abstract
{}
{We investigate whether the cylindrical (galactocentric) radial
  velocity gradient of $\sim-3\kmseckpc$, directed radially from the
  Galactic center and recently observed in the stars of the solar
  neighborhood with the \emph{RAVE} survey, can be explained by the
  resonant effects of the bar near the solar neighborhood.}
{We compared the results of test particle simulations of the Milky Way
  with a potential that includes a rotating bar with observations from
  the \emph{RAVE} survey. To this end we applied the \emph{RAVE}
  selection function to the simulations and convolved these with the
  characteristic \emph{RAVE} errors. We explored different ``solar
  neighborhoods" in the simulations, as well as different bar models.}
{We find that the bar induces a negative radial velocity gradient at
  every height from the Galactic plane, outside the outer Lindblad
  resonance and for angles from the long axis of the bar compatible
  with the current estimates. The selection function and errors do not
  wash away the gradient, but often make it steeper, especially near
  the Galactic plane, because this is where the \emph{RAVE} survey is
  less radially extended. No gradient in the vertical velocity is
  present in our simulations, from which we may conclude that this
  cannot be induced by the bar.}
{}

	
\maketitle

\section{Introduction}
Many of the past efforts in modeling the mass distribution of the
Milky Way have assumed that the Galaxy is axisymmetric and in a steady
state. However, there is a wealth of evidence that these assumptions
are not really valid. The two most important deviations from
axisymmetry are the spiral arms and the bar. These features are not
only apparent as non-axisymmetric density enhancements, but they also
have long-range gravitational effects. In particular, the bar modifies
the kinematics of the outer parts of the Galactic disks, far beyond
its extension, through resonant interactions.

That the velocity distribution of stars very near to the Sun is not
smooth (as one would expect in a steady state axisymmetric system),
but instead rich in substructures, has been established
observationally thanks to data from the \emph{Hipparcos} satellite and
other surveys (\citealt{Dehnen1998,Famaey2005,Antoja2008}). Several
authors have explained these substructures as being due to orbital
resonant effects of the bar (\citealt{Dehnen2000, Fux2001}), of the
spiral arms (\citealt{Mayor1970,DeSimone2004}; \citealt{Antoja2011}),
or both (\citealt{Antoja2009,Quillen2011}).

Using data from the \RAVE survey (\citealt{Steinmetz2006}),
\cite{Antoja2012} discovered that some of the kinematic substructures
detected in the vicinity of the Sun can be traced further, both on and
above the plane of the Galaxy, up to $\sim 0.7\Kpc$.

But \emph{RAVE} also made it possible to discover large scale
streaming motions. \cite{Siebert2011grad} (in the rest of the paper
S11) used a sample of $213,713$ stars to discover a gradient in the
mean galactocentric radial velocity that decreases outward with
Galactic radius. S11 show that this gradient was also present when
using only the $29,623$ red clump stars in their sample, whose distances
are more accurate. \cite{Siebert2012} modeled the gradient as caused
by a long-lived spiral pattern. \cite{Williams2013} (hereafter W13)
studied the 3D velocity distribution of red clump stars in \emph{RAVE}
in detail, confirmed the existence of the radial velocity gradient and
also discovered a more complicated vertical velocity distribution than
expected, attributing it to secular phenomena in the Galaxy.
\cite{Faure2014} generalized to 3D the model for the spiral arms
presented in \cite{Siebert2012}, which now also depends on the
distance from the Galactic plane. This model nicely predicts a
behavior for the mean vertical velocity that is similar to what is
observed in W13 (i.e., resembling ``rarefaction-compression'' waves),
together with the radial velocity gradient.

On the other hand, \cite{Monari2013} (hereafter M13), used 3D test
particle simulations to show that the gravitational effects of the bar
can significantly affect the kinematics of stars near the Sun, even at
distances from the Galactic plane up to at least $z\sim1\Kpc$ for the
thin disk and $z\sim2\Kpc$ for the thick disk. These results imply
that some of the substructures found in \cite{Antoja2012} could also
be caused by the bar.

In this paper we investigate an alternative explanation for the
observed radial velocity gradient, beyond that caused by the spiral
arms, by suggesting that it can be created by the bar. To do so, we
compare the results of the test particle simulations in M13 with the
\emph{RAVE} data.

The paper is organized as follows. In Sect.~\ref{sect:sim} we
summarize the salient characteristics of the simulations from M13, and
in Sect.~\ref{sect:sf} we describe how we apply the \RAVE selection
function and error convolution to them, to mimic a \RAVE catalog. In
Sect.~\ref{sect:res} we present the results. In
Sect.~\ref{sect:expl} we explain how the bar can create a radial
velocity gradient as observed. In Sect.~\ref{sect:disc} we discuss
the similarities between our results and the ones in W13 and in
Sect.~\ref{sect:concl} we conclude.

\section{Simulations}\label{sect:sim}
We use the 3D test particle simulations of the thin and thick disk of
the Milky Way described in M13. The rigid background gravitational
potential in these simulations includes an axisymmetric part (composed
of a dark halo, and a thin and a thick disk) and a non-axisymmetric
perturbation to represent the Galactic bar. The bar potential follows
a \cite{Ferrers1870} model and we varied its structural parameters
(with values taken from the literature) but with a constant pattern
speed $\Omegab=50\kmseckpc$. For the comparison with the \RAVE data we
focus on simulations with three bar potentials: the default bar, the
long bar, and the less massive bar (corresponding to GB2, LB2, and GB1
in M13). We choose the snapshot of the simulations at $t=24\Tbar$,
which corresponds to $\sim 3\Gyr$ after the introduction of the bar in
the simulations (see M13). In the default bar case, there are
$\Nthin=10^9$ particles in the thin disk population and
$\Nthick=2\times10^8$ particles in the thick disk (to have a
thick-to-thin density of particles normalization of $\sim 10\%$ at the
Sun). For the remaining cases we only have low resolution simulations
with $\Nthin=5\times10^7$ and $\Nthick=10^7$ particles for the thin
and the thick disk, respectively. The parameters of the simulations
are summarized in Table~\ref{tab:params}.

In this paper $\left(R,\phi,z\right)$ are the Galactocentric
cylindrical coordinates, where $\phi$ is the angle from the long axis
of the bar in the direction of rotation of the Galaxy. The cylindrical
velocities are
$\left(v_R,v_\phi,v_z\right)=\left(\dot{R},R\dot{\phi},\dot{z}\right)$.
Right ascension, declination, and heliocentric distance are denoted as
$\left(\alpha,\delta,d\right)$, and the proper motions and line of
sight velocity as $\left(\mu_\alpha,\mu_\delta,\vlos\right)$.

\begin{table}
  \caption{Parameters of the bar and location of the main resonances.}
  \label{tab:params}
  \centering
  \begin{center}
    \begin{tabular}{@{\extracolsep{-5pt}}llll}\hline\hline
      Parameter & Default bar & Long bar & Less massive bar\\ \hline
      $\Mbar(\Msun)$ & $2\times10^{10}$ & $2\times10^{10}$ & $10^{10}$
      \\ $a(\mathrm{kpc})$ & $3.5$ & $3.9$ & $3.5$ \\ $b(\mathrm{kpc})$ & $1.4$ &
      $0.6$ & $1.4$ \\ $c(\mathrm{kpc})$ & $1.0$ & $0.1$ & $1.0$
      \\ $\RCR(\mathrm{kpc})$ & $4.91$ & $4.94$ & $4.54$ \\ $\ROLR(\mathrm{kpc})$ &
      $7.69$ & $7.69$ & $7.40$ \\ \hline
    \end{tabular}
  \end{center}
\end{table}

\section{Selection function and errors}\label{sect:sf}

\subsection{Red clump stars}
Both S11 and W13 use red clump stars because they are promising
standard candles, since they are easy to identify in the HR diagram,
and while being relatively unaffected by extinction their $K$-band
magnitude depends only weakly on metallicity and age. In W13 the red
clump is selected from the internal \emph{RAVE} release from October
2011\footnote{Here $K$ is used to denote $K$-band magnitudes in the
  $2MASS$ system, while $\JBB$ and $\KBB$ denote the $J$ and $K$-band
  magnitudes in the \cite{BesselBrett1989} system.} (see the DR3
paper, \citealt{Siebert2011b}, for stellar parameter determination),
as those stars with
\begin{equation}
  0.55 \leq \JBB-\KBB \leq 0.8 \quad \mathrm{and} \quad 1.8 \leq \log
  g \leq 3.0.
\end{equation}
The absolute magnitude associated with the red clump stars is taken to
be $\MK=-1.65$, following \cite{Alves2000}.

\subsection{RAVE selection function}
We applied the \emph{RAVE} selection function in the space of
observables $\left(\alpha,\delta,K\right)$ to the simulations. We did
this as follows:
\begin{itemize}
\item we chose the position of the Sun in the simulated Galaxy; the
  default position is
  $\left(R,\phi,z\right)=\left(8\Kpc,-20\degree,0\right)$, in the
  range of current determinations (\citealt{BissantzGerhard2002});
\item we transformed the spatial coordinates of the simulation into
  the observables $\left(\alpha,\delta\right)$ and heliocentric
  distance $d$;
\item we assigned $K$ magnitudes to the particles in the simulation,
  assuming they are red clump stars with $\MK=-1.65$. Then the
  apparent magnitude is given by $K=\MK+5\left(\log_{10}d-1\right)$;
\item we binned the red clump stars used in W13 in the
  $(\alpha,\delta,K)$ space, with $\Nb$ bins of size $10\degree \times
  10\degree \times 0.2\mg$ and $\alpha \in [0,360\degree]$, $\delta
  \in [-90\degree,0]$, $K \in [2,12]$;
\item we binned in the same space and in the same way the particles
  from the simulation that are inside a sphere of radius $3\Kpc$ from
  the Sun;
\item for $i=1,...,\Nb$, if the $i$-th bin in the simulation contains
  $N_i \ge N^R_i$ particles (where $N^R_i$ is the number of stars in
  the same bin in \RAVE) we randomly downsampled it to $N^R_i$
  particles; if $N^R_i > N_i \ge N^R_i - \sqrt{N^R_i}$ we kept the
  $N_i$ particles (because $N^R_i$ and $N_i$ differ less than the
  corresponding Poisson noise error); however, not all the bins of the
  simulations are populated enough: we excluded from the comparison
  those bins with $N_i<N^R_i - \sqrt{N^R_i}$ (this only happened for
  less than $\sim 0.05\%$ of the bins for the high resolution
  simulations, and also in the low resolution case after the treatment
  explained in Sect.~\ref{sect:oth_b_models}).
\end{itemize}  
We repeated this procedure obtaining $100$ different random samples of
each simulation, where the $\alpha$, $\delta$ and $K$ distribution are
almost perfectly matched and the total number of star particles
differs from the \RAVE red clump sample in W13 by less than $0.5 \%$
for the high resolution simulations, leaving $72,064$ particles in the
default bar case.
\begin{figure}
  \centering
  \includegraphics[width=\columnwidth]{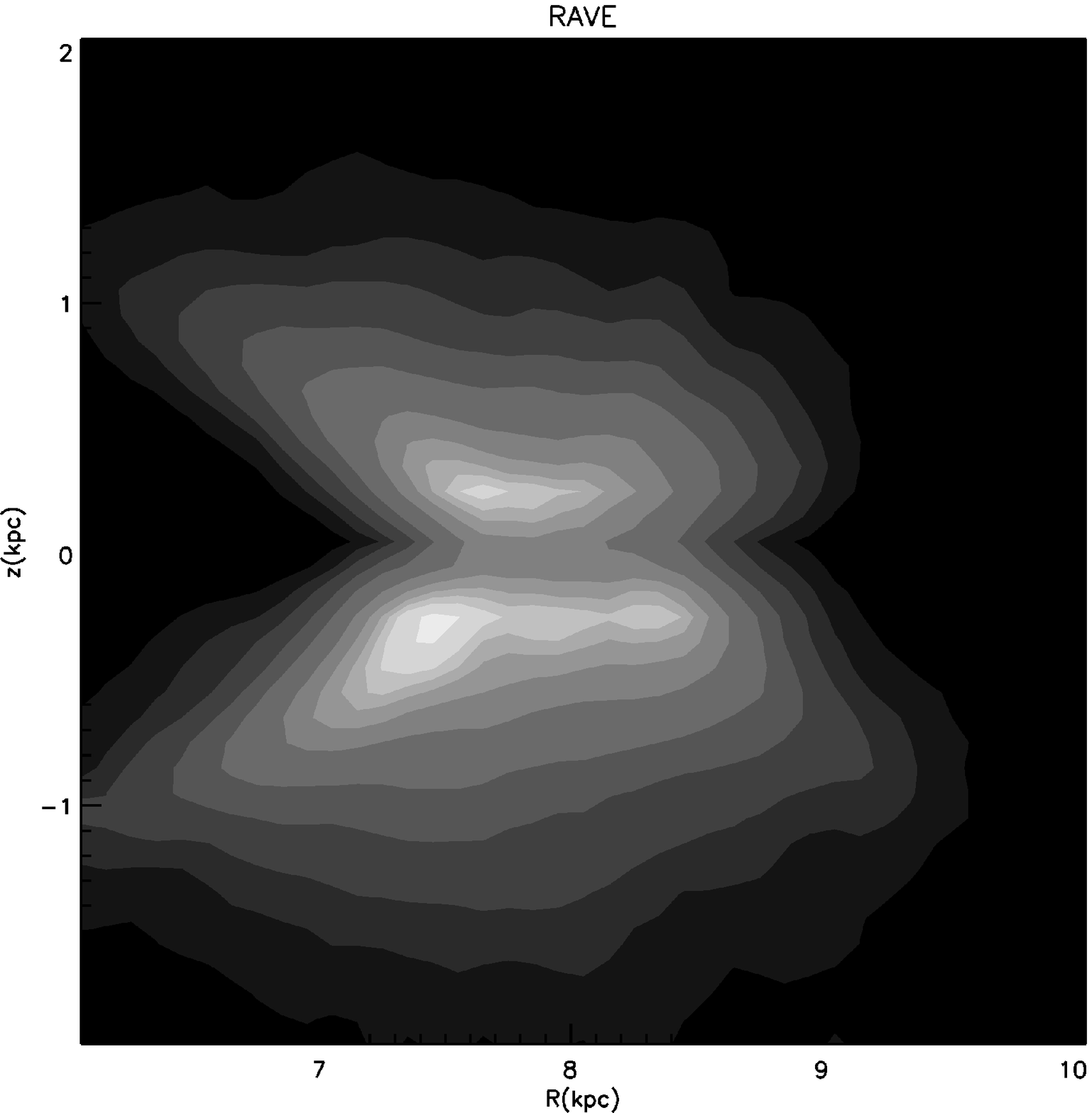}
  \includegraphics[width=\columnwidth]{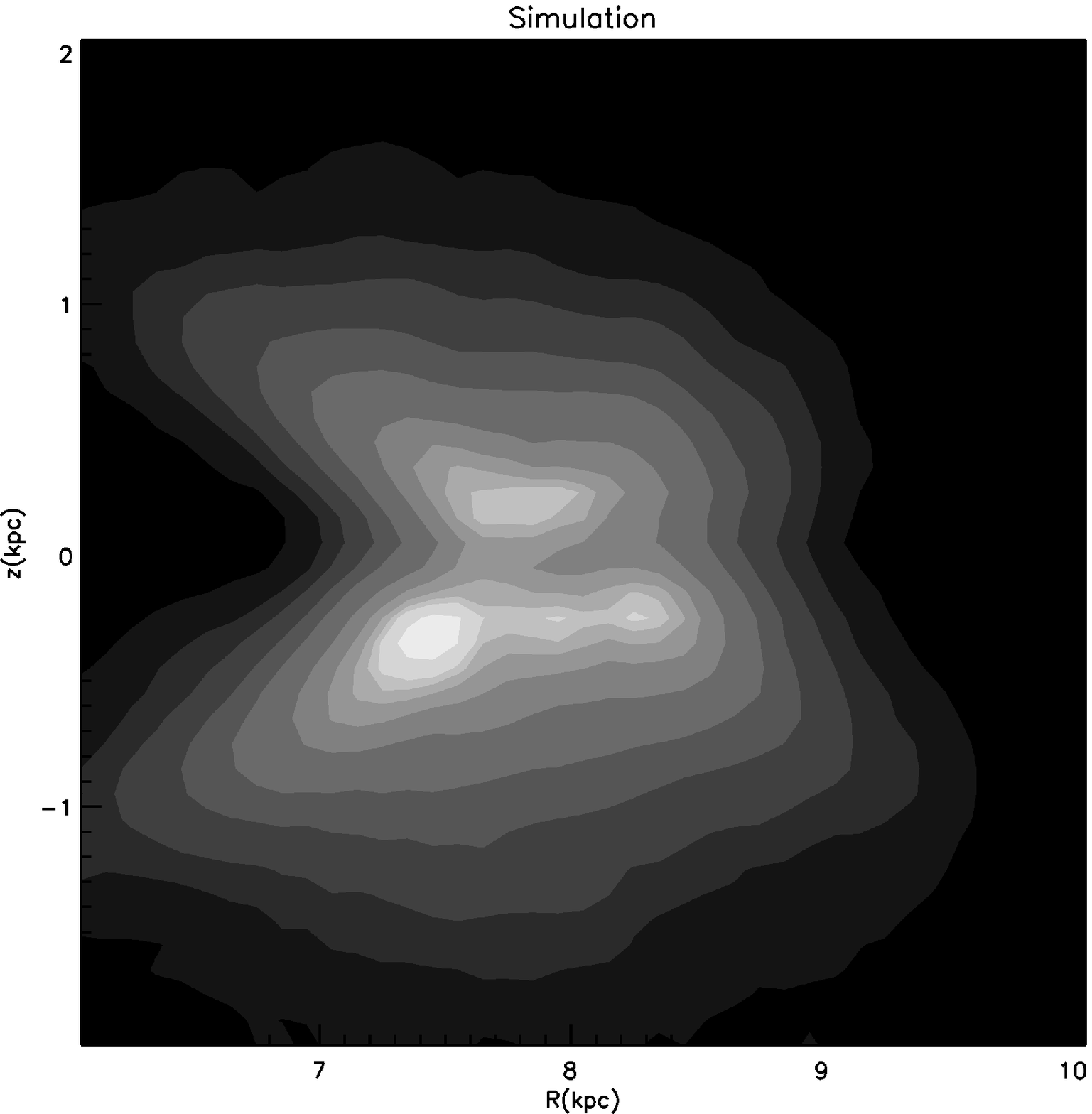}
  \caption{$(R,z)$ distribution of stars in the \emph{RAVE} sample
    used in W13 (top) and in the simulation with default Sun's
    position after the application of the \RAVE selection function
    (bottom). The contours enclose $2,6,12,21,33,50,68,80,90,95$ and
    $99\%$ of the stars.}
  \label{fig:Rz_SF}
\end{figure}
The top panel of Fig.~\ref{fig:Rz_SF} shows the $(R,z)$ distribution
of red clump stars in W13, while in the bottom we have plotted the
result of the procedure described above for our standard simulation
and Sun's position. We see that we are successful in reproducing
how the different \RAVE fields are populated. The differences in
$(R,z)$ (e.g., inside the contour enclosing $21\%$ of the stars) are
due to the small differences described above in the bins in the
$(\alpha,\delta,K)$ space.

\subsection{Error convolution}
For the comparison between data and simulations we proceeded to
convolve the latter with the errors estimated for the \RAVE survey.
 
We produced a simple error model, where the errors in proper motion
and line of sight velocity are function of the $K$ magnitude
only. This was done by fitting second order polynomials to $K$
vs. $\epm$ and $K$ vs. $\elos$ for the red clump stars in DR3. We also
derived an error in distance $e_d$ propagating the error in $K$
($e_K\sim 0.04\mg$) and the spread in absolute magnitudes of the red
clump ($e_{\MK}=0.22\mg$, \citealt{Alves2000}). The resulting relative
error in distance is $e_d/d\sim0.1$, that we assumed to be the same
for all particles in our simulations.

The error convolution was done in the observable space
$(\alpha,\delta,d,\mu_\alpha,\mu_\delta,\vlos)$, assuming Gaussian
errors in each quantity.

\section{Results}\label{sect:res}
\subsection{Default case}
As a default case we place the Sun at
$(R,\phi,z)=(8\Kpc,-20\degree,0)$ and we choose the default bar
model. In this case $R_0/\ROLR=1.04$, where $R_0$ is $R$ of the Sun
and $\ROLR$ is the Galactocentric distance of the outer Lindblad
resonance.
\begin{figure*}
  \centering
  \includegraphics[width=0.3\textwidth]{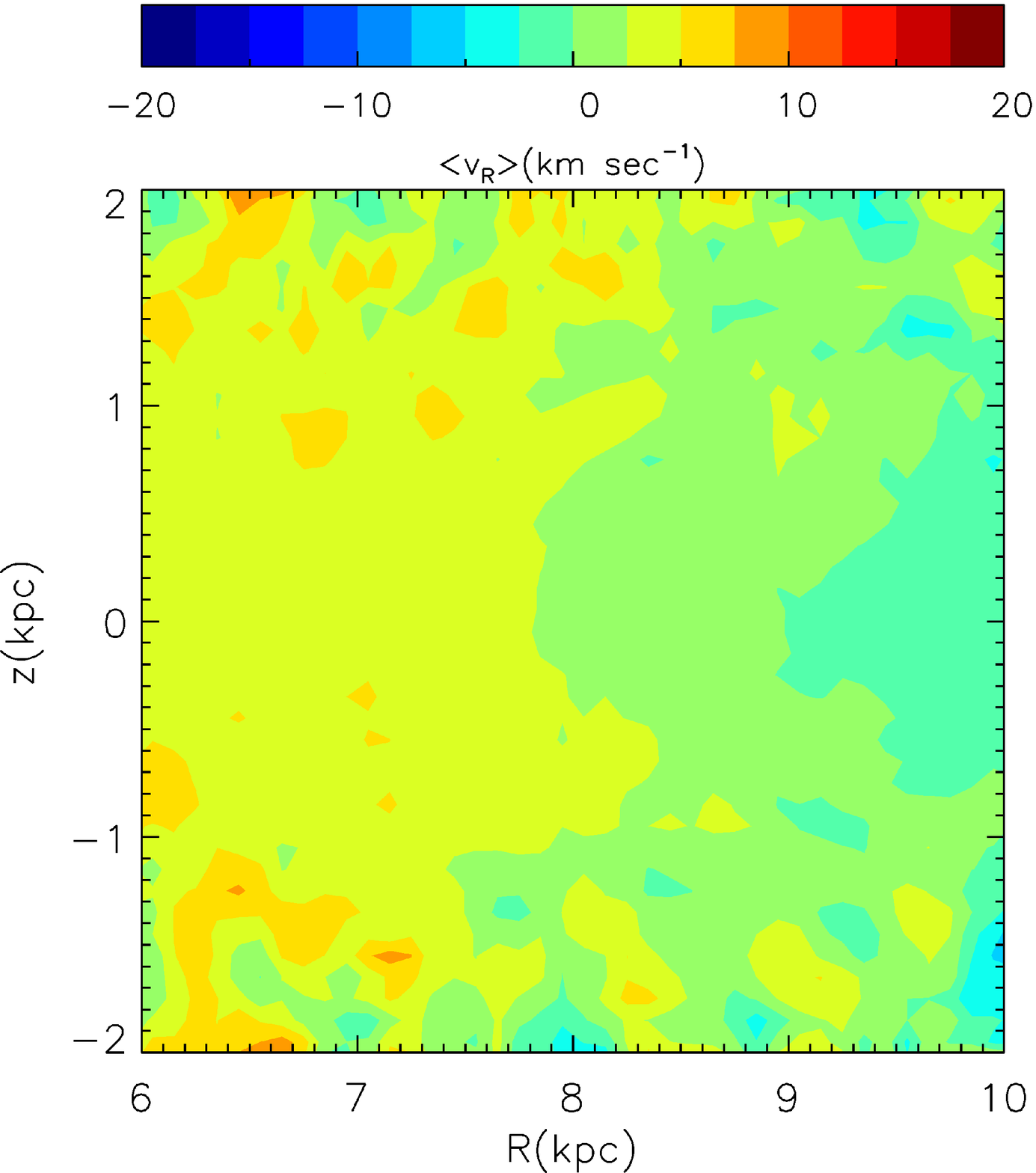}
  \includegraphics[width=0.3\textwidth]{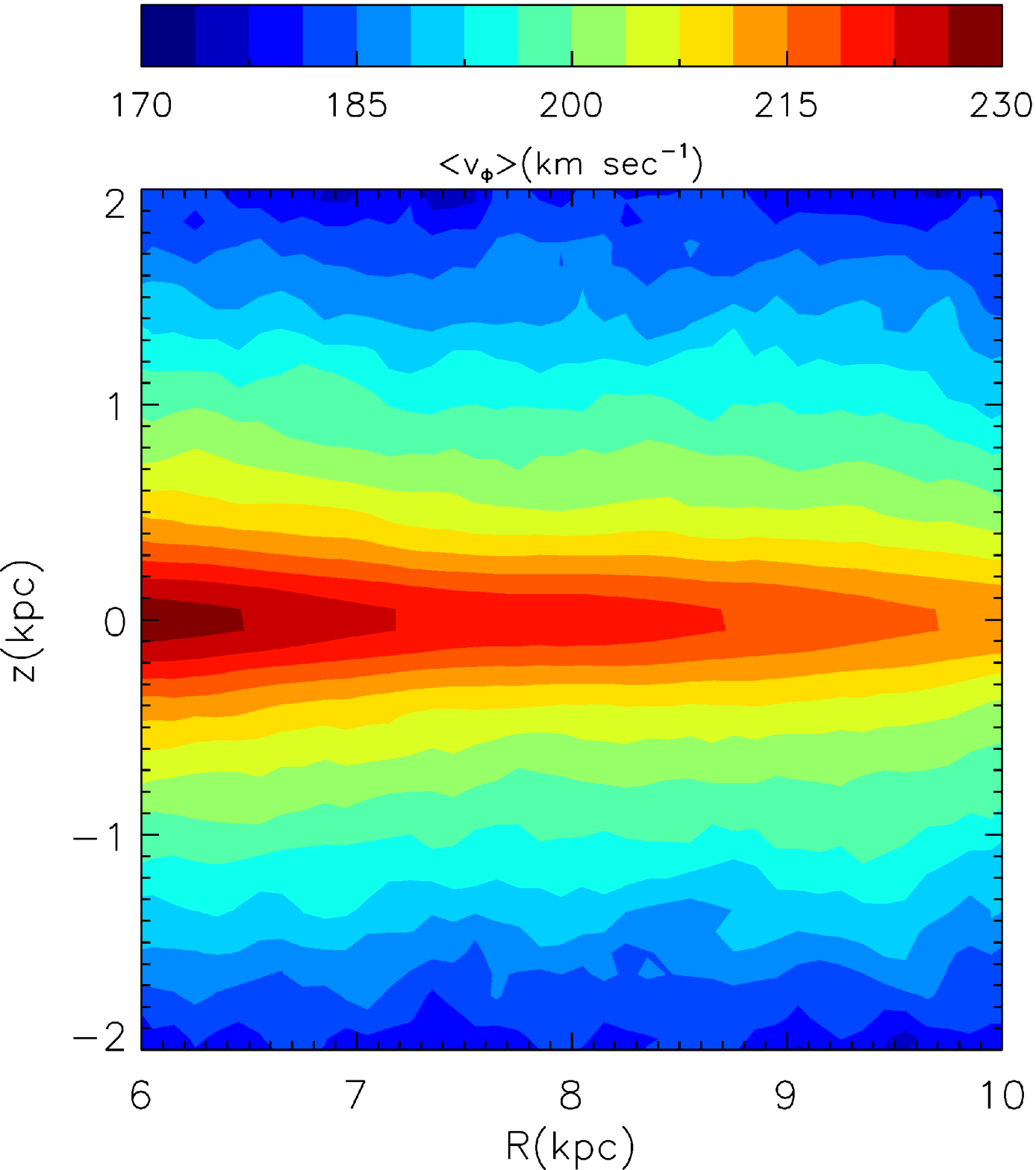}
  \includegraphics[width=0.3\textwidth]{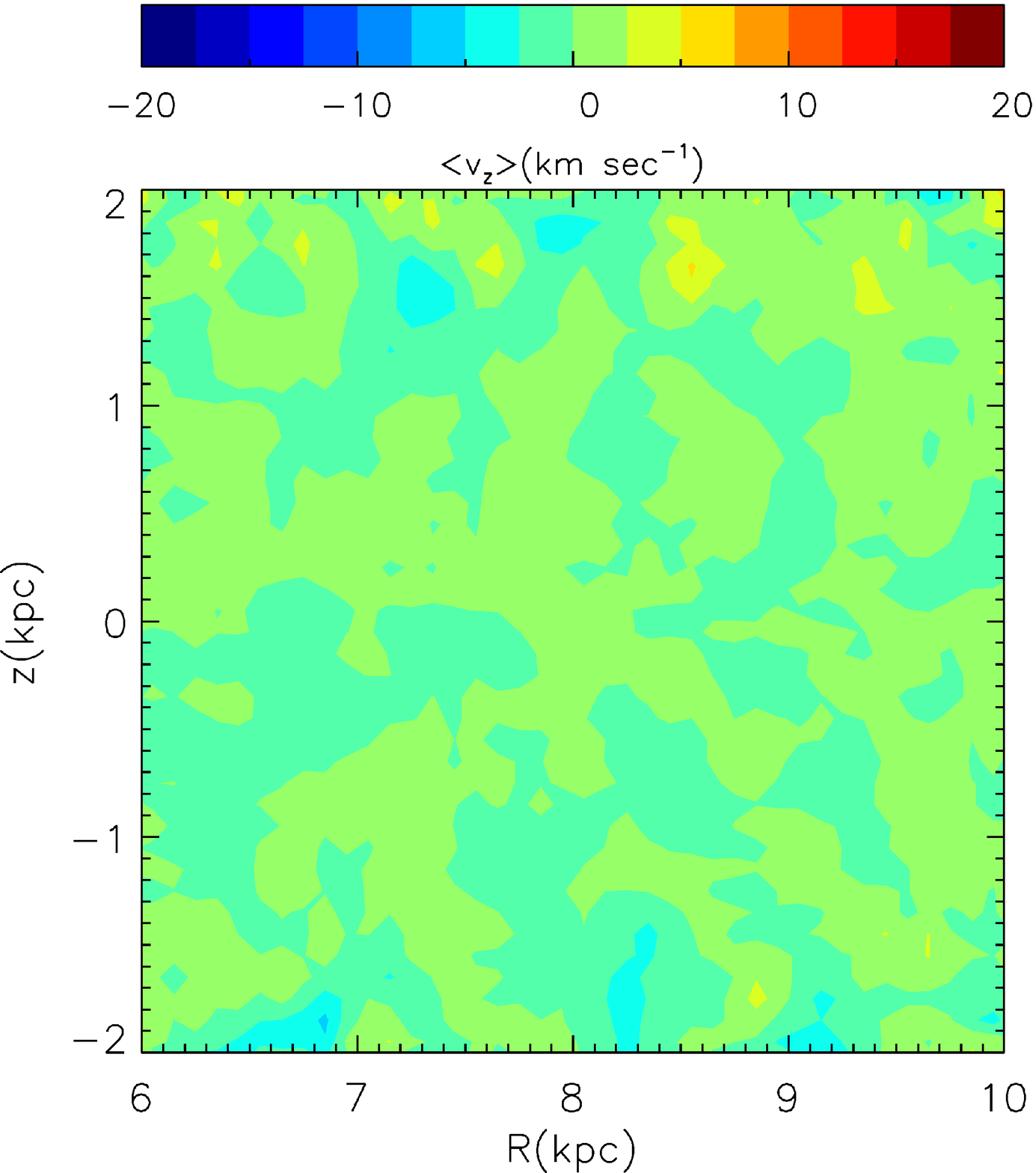}
  \includegraphics[width=0.3\textwidth]{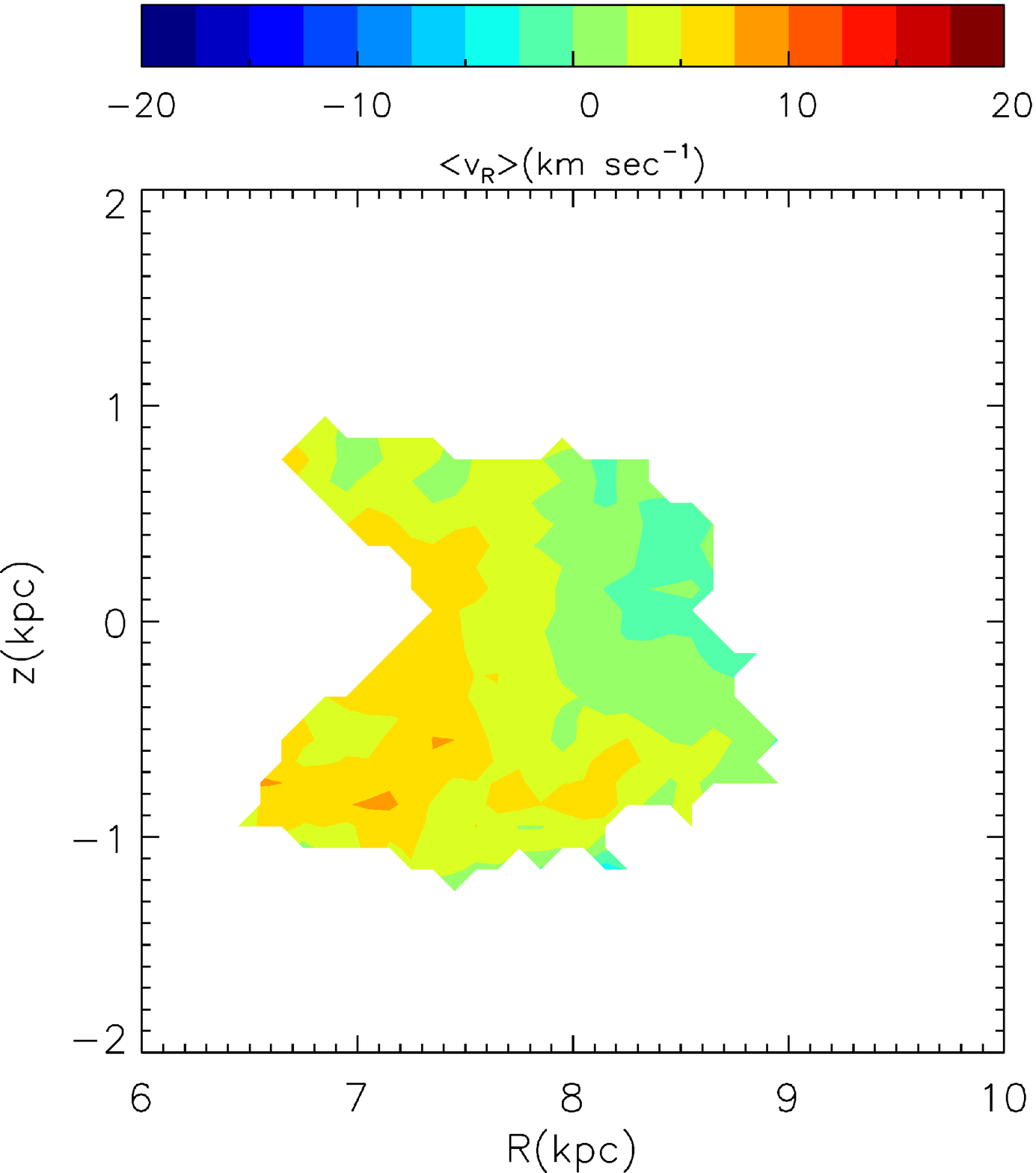}
  \includegraphics[width=0.3\textwidth]{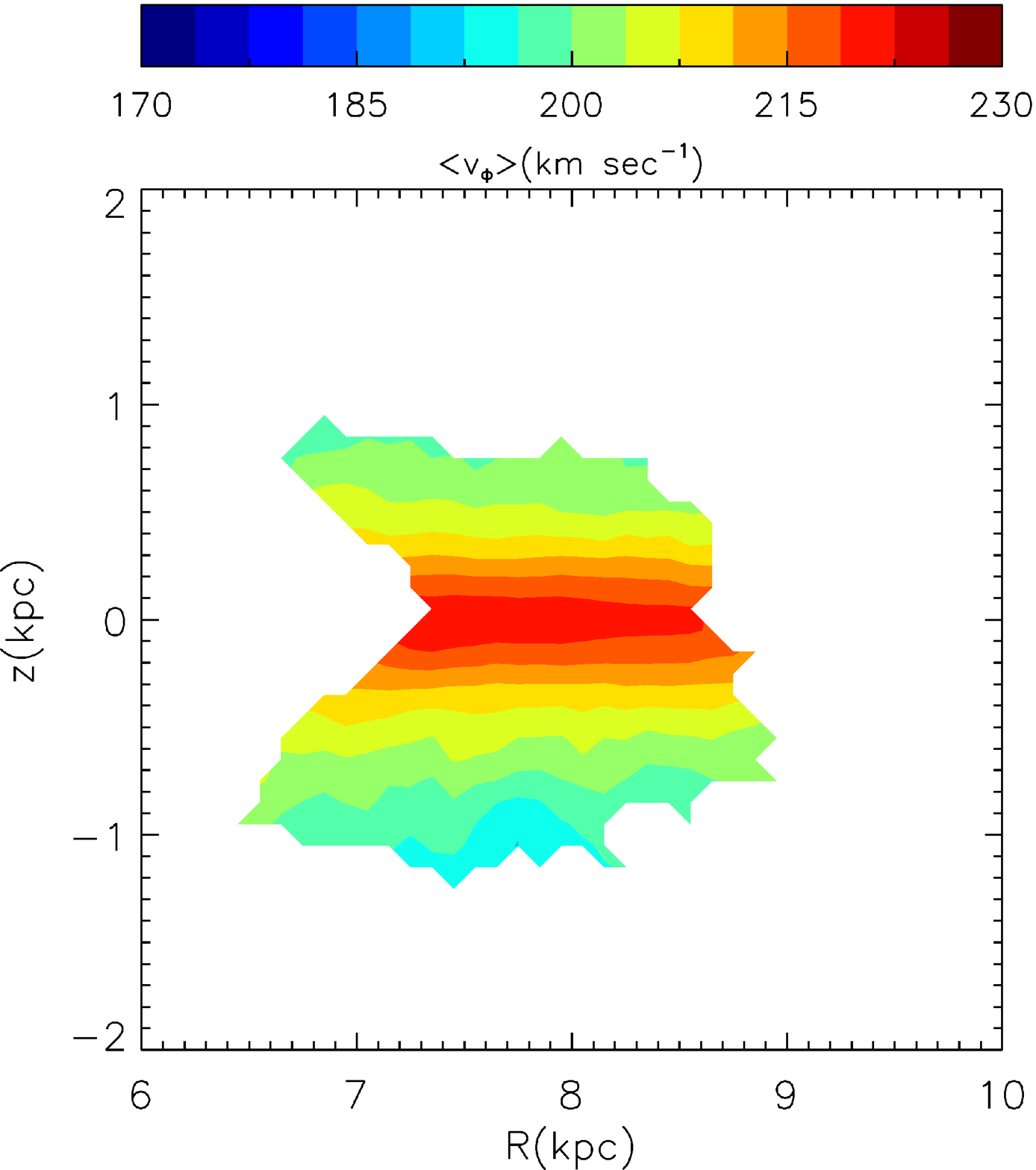}
  \includegraphics[width=0.3\textwidth]{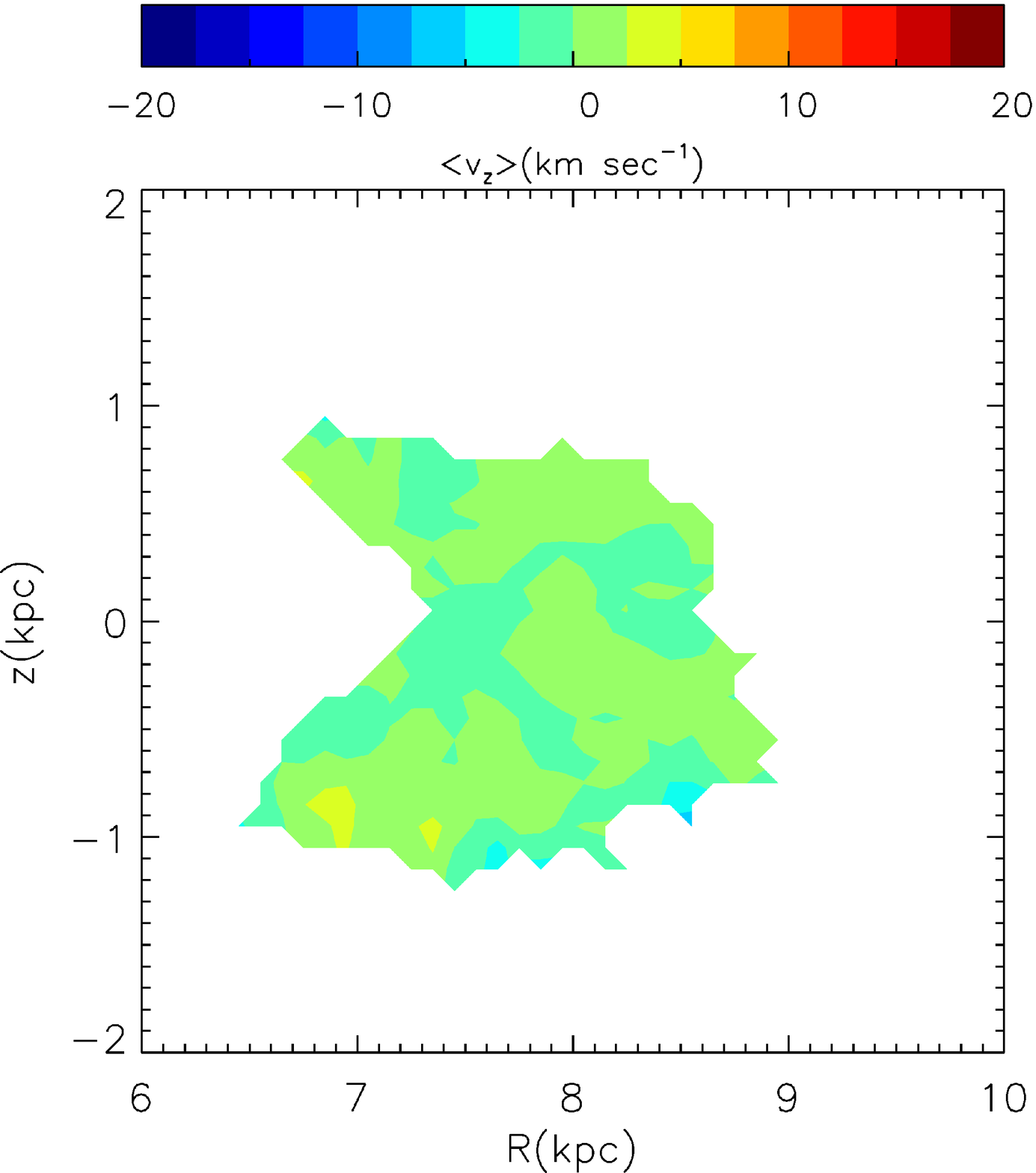}
  \caption{Average velocities for particles inside a sphere of radius
    $3\Kpc$, when the Sun is placed at
    $(R,\phi,z)=(8\Kpc,-20\degree,0)$ and for the default bar
    model. Top row: contour plots in the $(R,z)$ plane, with binsize
    $100\pc$ and box smoothing on a scale of $200\pc$. Bottom row: as
    in the top row, but after the application of the \emph{RAVE}
    selection function. The averages in the bottom panels correspond
    to those obtained using $100$ random samples of the simulation.}
  \label{fig:GB_8kpc_20deg_all}
\end{figure*}

The top row of Fig.~\ref{fig:GB_8kpc_20deg_all} shows the average
velocities as a function of $R$ and $z$, for all the particles inside
a sphere of radius $3\Kpc$ from the Sun, with no error convolution yet
applied. From left to right we show contour plots of $\avvR$,
$\avvphi$, and $\avvz$. As in W13, the data are averaged inside bins
of $100\pc$ size in $(X',Y')$, box smoothed on a scale of $200\pc$.
As it is apparent from these plots, $\avvR$ decreases increasing the
$R$ distance in the simulated Galaxy. Moreover, $\avvR$ is rather
symmetric with $z$. The central panel shows that the rotational
velocity of the stars in the Galactic disk(s) $\avvphi$ decreases with
distance from the plane. This is because the velocity dispersion
increases with $z$ and because the asymmetric drift increases with the
velocity dispersion (\citealt{BT2008}). The rightmost panel shows
instead how $\avvz\sim0$ everywhere in the simulated sample, i.e., the
distribution function of our simulations is an even function of $v_z$.

The bottom row of Fig.~\ref{fig:GB_8kpc_20deg_all} shows the same
quantities as in the top panels, after the application of the
\emph{RAVE} selection function. The plotted values were obtained
averaging over the $100$ random samples of the simulation, distributed
in $(\alpha,\delta,K)$ as in \emph{RAVE}. We only consider the bins
including more than $50$ particles. These contour plots show that the
decreasing $\avvR$ gradient is preserved after the selection function
has been applied to the simulation. In fact the gradient is even
enhanced: the yellow regions at $R\sim7\Kpc$ are formed by particles
with $\avvR>5\kmsec$, and the blue/green regions at $R\sim 8.5\Kpc$ by
particles with slightly negative $\avvR$. From the second and third
panel we also see that the selection function does not induce any
significant difference in $\avvphi$ and $\avvz$: unlike W13, in the
samples presented in this work we do not detect any significant
$\avvz$ gradient with respect to $z$ or $R$.

\begin{figure}
  \centering
  \includegraphics[width=\columnwidth]{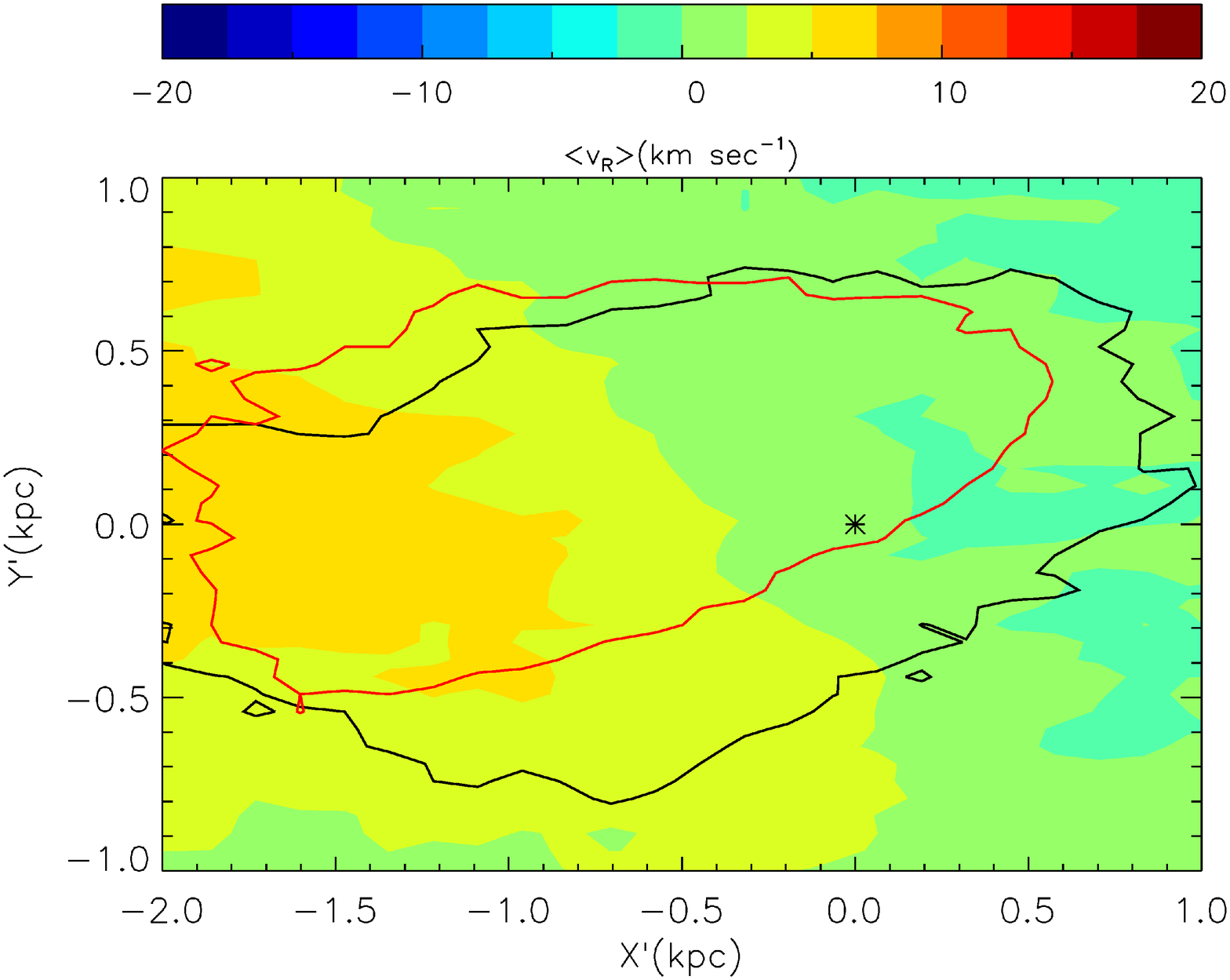}
  \caption{Trends in $\avvR$ as a function of the cartesian
    coordinates $(X',Y')$ centered on the Sun, with the Galactic
    center at $(X',Y')=(-8\Kpc,0)$. These are computed for particles
    inside a sphere of radius $3\Kpc$, when the Sun is placed at
    $(R,\phi,z)=(8\Kpc,-20\degree,0)$ and for the default bar model.
    The plotted data are averaged inside bins of $100\pc$ size in
    $(X',Y')$, box smoothed on a scale of $200\pc$. The contours
    enclose $90\%$ of particles with $-1.5\Kpc<z<0$ (black) and
    $0<z<1.5\Kpc$ (red), when the \RAVE selection function is
    applied.}\label{fig:xy}
\end{figure}

The reason why the selection function enhances the $v_R$ gradient is
readily understood from Fig.~\ref{fig:xy}.
Here $X'$ and $Y'$ are the cartesian coordinates centered at the Sun, the
Galactic Center is placed at $(X',Y')=(-8\Kpc,0)$, and the colors
represent $\avvR$ for particles inside a sphere of radius $3\Kpc$ from
the Sun in bins of $100\pc$ size. The contours enclose $90\%$ of
particles with $-1.5\Kpc<z<0$ (black) and $0<z<1.5\Kpc$ (red), when
the \RAVE selection function is applied. These contours therefore show
that the selection function encloses mostly particles with negative
$X'$ and positive $Y'$ (with $\phi<-20\degree$), where the gradient is
steeper.

This analysis shows that $v_R$ is the velocity component most
influenced by the bar (and that no signature is readily apparent in
$v_z$) and therefore we focus in the rest of the paper on the $R$
gradients of $\avvR$ and on their dependence on $z$.

\begin{figure*}
  \centering
  \includegraphics[width=\textwidth]{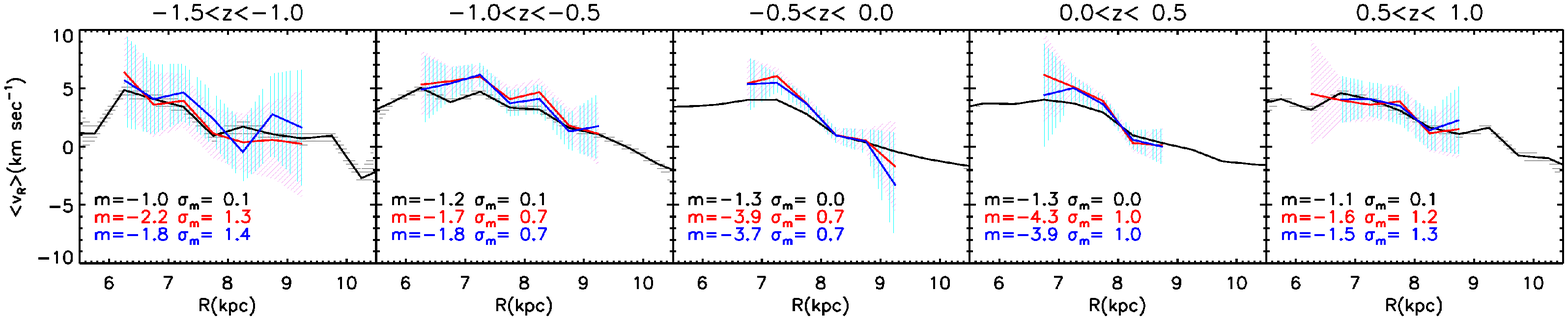}
  \caption{Average cylindrical radial velocity $\avvR$ as a function
    of radial distance in the plane $R$, for particles inside a sphere
    of radius $3\Kpc$, when the Sun is placed at
    $(R,\phi,z)=(8\Kpc,-20\degree,0)$ and for the default bar
    model. The bin size is $0.5\Kpc$. Each panel corresponds to
    particles in a certain range of $z$. The black curve corresponds
    to the whole sample and the error bands are the statistical error
    on the mean. The other two curves represent the sample after the
    application of the \emph{RAVE} selection function, without (red
    line) and with (blue line) error convolution. The quantities
    represented by the red and blue curves are averaged over $100$
    random subsamples of the simulation. The error bands correspond to
    the maximum error on the mean amongst the different samples. We
    only show the bins with errors smaller than $5 \kmsec$ and more
    than $50$ particles.}
  \label{fig:trends_GB_8kpc_20deg}
\end{figure*}
\begin{figure*}
  \centering
  \includegraphics[width=\textwidth]{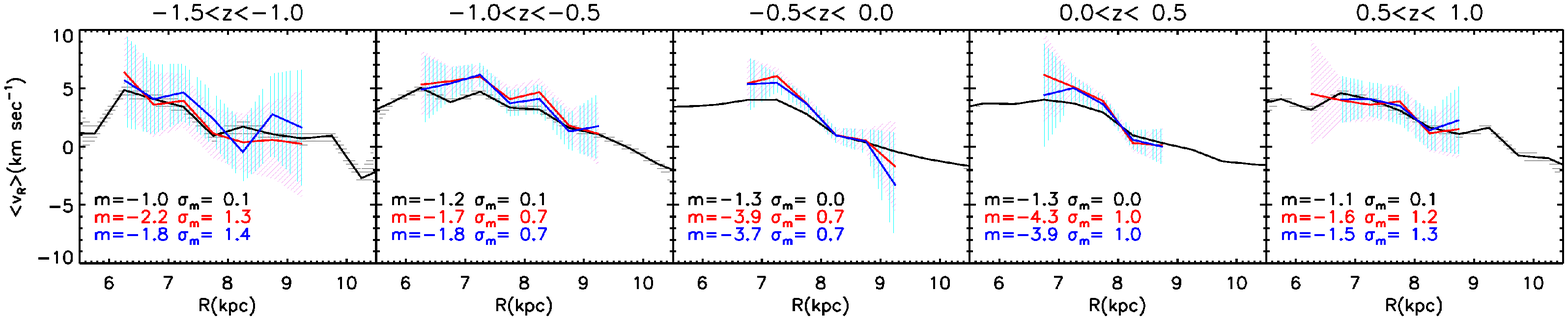}
  \caption{As in Fig.~\ref{fig:trends_GB_8kpc_20deg}, but with the Sun
    placed at $(R,\phi,z)=(8\Kpc,-40\degree,0)$.}
  \label{fig:trends_GB_8kpc_40deg}
\end{figure*}
\begin{figure*}
  \centering
  \includegraphics[width=\textwidth]{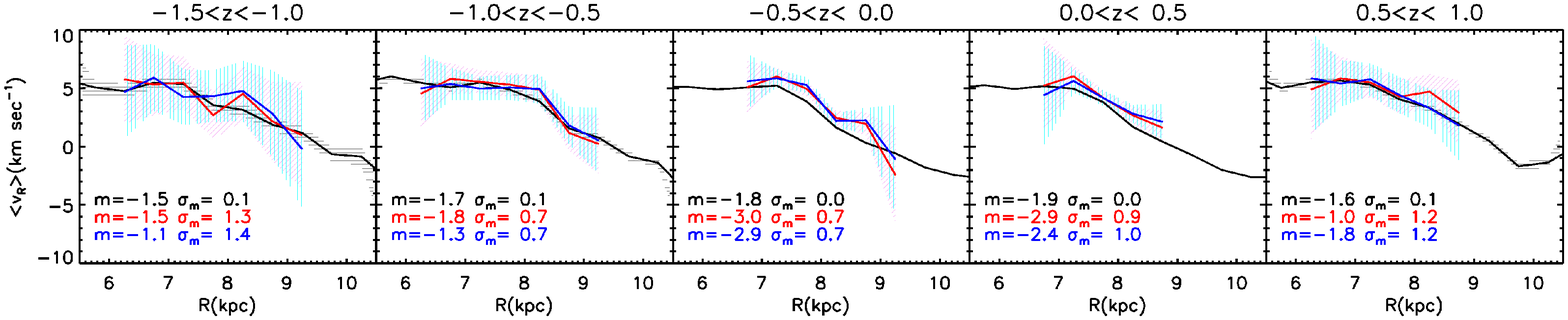}
  \caption{As in Fig.~\ref{fig:trends_GB_8kpc_20deg}, but with the Sun
    placed at $(R,\phi,z)=(9\Kpc,-20\degree,0)$.}
  \label{fig:trends_GB_9kpc_20deg}
\end{figure*}
\begin{figure*}
  \centering
  \includegraphics[width=\textwidth]{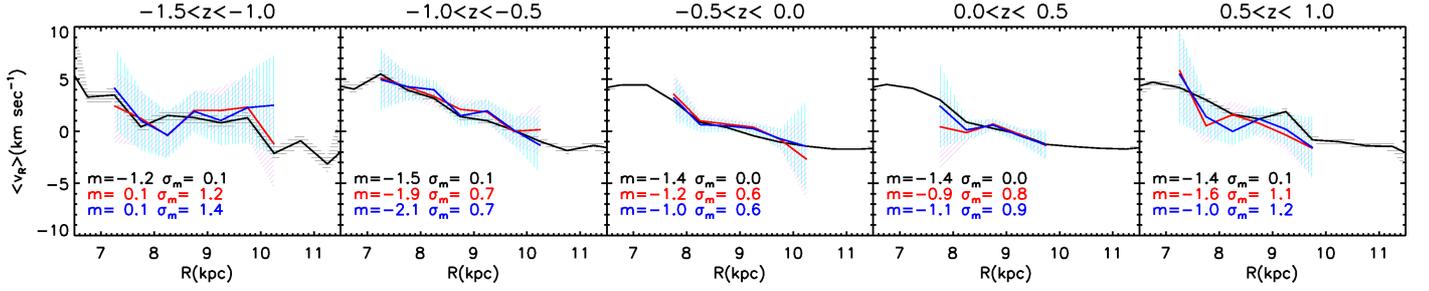}
  \caption{As in Fig.~\ref{fig:trends_GB_8kpc_20deg}, but with the
    long bar. Note that in this case, the simulation has been mirrored
    with respect to the $z=0$ plane, which implies that the black
    curves in the 2nd and 5th, and in the 3rd and 4th panels are
    identical. However, the \RAVE selection function does depend on
    Galactic latitude, resulting in different blue and red curves in
    each panel.}
  \label{fig:trends_LB_8kpc_20deg}
\end{figure*}
\begin{figure*}
  \centering
  \includegraphics[width=\textwidth]{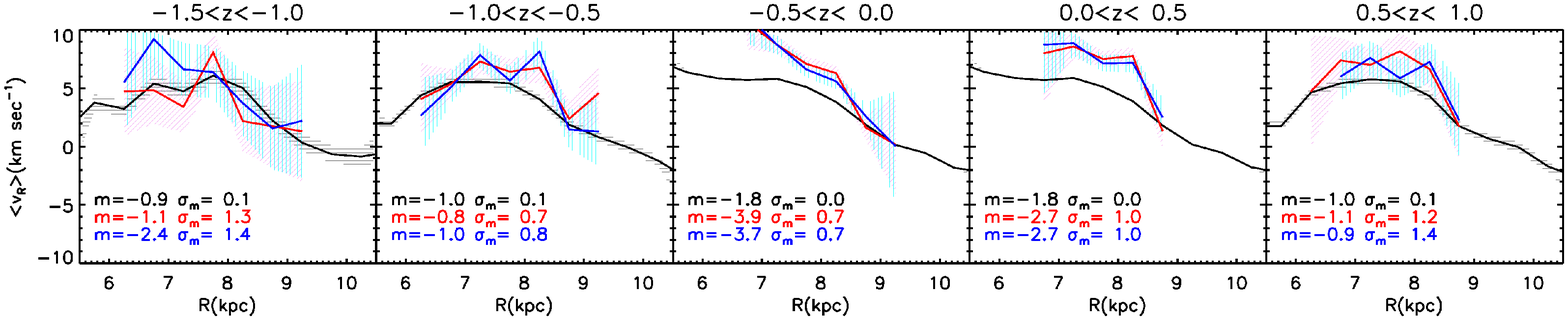}
  \caption{As in Fig.~\ref{fig:trends_GB_8kpc_20deg}, but with less
    massive bar. As in the previous figure, the simulations have been
    mirrored with respect to the $z=0$ plane.}
  \label{fig:trends_LMB_8kpc_20deg}
\end{figure*}

In Fig.~\ref{fig:trends_GB_8kpc_20deg} we look closer at the $\avvR$
trends with $R$, slicing the particles at different $z$. The quantity
$\avvR$ is computed inside $R$ bins of size $0.5\Kpc$. The black line
has been obtained using all the particles in the simulations inside a
sphere of $3\Kpc$ radius from the Sun, the red line those that remain
after applying the \emph{RAVE} selection function only, and the blue
line the case where the error convolution is applied together with the
selection function. The shaded areas represent the standard errors of
the mean inside each bin. For the red and blue curves we show the
maximum error amongst the $100$ random samples\footnote{We could not
  use the standard deviation of these 100 random samples, as they are
  not completely independent. This is because in some of the
  $\left(\alpha, \delta, K\right)$ bins there are as many particles in
  the simulations as stars in the \RAVE red clump sample.}. We only
show the bins with errors smaller than $5\kmsec$ and including more
than $50$ particles. The black line confirms what we saw in
Fig.~\ref{fig:GB_8kpc_20deg_all}, namely that $\avvR$ decreases with
$R$ in each of the $5$ slices in $z$.

The numbers in the bottom right of each panel quantify the magnitude
of the gradient: $m$ represents the slope of the linear regression of
the $\avvR$ values inside the plotted bins, and $\sigma_m$ its
uncertainty (computed from the errors represented by the shaded
areas). Although $\avvR$ slowly decreases ($m\sim-1\kmseckpc$) with
$R$ in each $z$ slice, the trend is not simple. It is the composition
of a flat/increasing gradient for $R\lesssim 7.5\Kpc$ and a decreasing
one for $R\gtrsim7.5\Kpc$. This break happens almost in correspondence
with the outer Lindblad resonance ($\ROLR=7.69\Kpc$). This composite
behavior makes $m$ shallower than if only the data points with
$R\gtrsim7.5\Kpc$ were to be fitted.

The \emph{RAVE} selection function (red line) in this default bar case
makes the gradients steeper, because the bins with $R\lesssim7.5\Kpc$
are excluded (or less populated) and for this reason they do not
reduce the slope. This is especially clear for the two slices
$-0.5\Kpc<z<0$ and $0<z<0.5\Kpc$, where the retained bins are all very
near to the Sun (because the \emph{RAVE} fields have $|b|>25\degree$,
see Fig.~\ref{fig:Rz_SF}). Since the Sun happens to be placed in the
middle of the region where $\avvR$ decreases and we only have the
nearest bins, the resulting gradient is very steep
($m\sim-4\kmseckpc$). This effect is mitigated further away from the
plane, because the $R$ extent of the survey becomes larger.

Finally, we note that the effect of error convolution is very small,
with the blue curves within the red (error-free) uncertainty bands.

\subsection{Other locations in the default bar}

\subsubsection{$R=8\Kpc$, $\phi=-40\degree$}

We consider now the effect of placing the Sun at a different position,
namely $(R,\phi,z)=(8\Kpc,-40\degree,0)$, thus at a larger angle from the
long axis of the bar.

Fig.~\ref{fig:trends_GB_8kpc_40deg} shows that, in this case, the
$v_R$ gradient in the whole sphere is steeper than in the default
case. In fact, the steepest gradient in the simulation is reached at
$\phi=-45\degree$ (the gradient is a periodic function of $\phi$, with
period $\pi$; for a detailed analysis of the periodic response of a
stellar disk to a bar see \citealt{Muhlbauer2003}). We see that in the
slices closest to the Galactic plane ($|z|<0.5\Kpc$) the effect of the
selection function is similar to that of the default case: the Sun is
placed in the region where the gradient is steepest and, since we
remove the outermost bins, the mean velocity gradient is strongly
enhanced. Again, at $|z|>0.5\Kpc$ the slope is smaller, because of the
larger $R$ extent of the sample.

\subsubsection{$R=9\Kpc$, $\phi=-20\degree$}
In Fig.~\ref{fig:trends_GB_9kpc_20deg} we present the analysis of the
radial velocity gradient behavior in a volume further out in the
Galaxy, namely at $(R,\phi,z)=(9\Kpc,-20\degree,0)$. This case was
chosen to illustrate the effect of the distance from the outer
Lindblad resonance, here $R_0/\ROLR=1.17$.

In this case the slope over the whole sphere is steeper than in the
default case. This happens because the volume is beyond the outer
Lindblad resonance and is therefore less affected by the particles
with $R\lesssim 7.5\Kpc$ (which were responsible for the positive/flat
gradient). However, the local gradient at $R=9\Kpc$ is shallower than
the one at $R=8\Kpc$, i.e., $\left|\frac{\de\avvR}{\de
  R}\left(9\Kpc\right)\right|<\left|\frac{\de\avvR}{\de
  R}\left(8\Kpc\right)\right|$. For this reason, when we apply the
selection function, for $|z|<0.5\Kpc$ the slope $m$ becomes smaller in
magnitude (than without selection function and than the default case
with selection function). On the contrary, for the slice with
$-1\Kpc<z<-0.5\Kpc$ it becomes steeper because in this case the
steepest part of the curve is included in the regression. Finally, in
the outermost slice ($-1.5\Kpc<z<-1\Kpc$) the low number of particles
make the $\avvR$ profile noisy and this washes out the $\avvR$
gradient.

\subsection{Other bar models}\label{sect:oth_b_models}
The simulations in M13 include two other bar models: the long bar
model with a different geometry but with the same mass as the default
bar, and a second one with same geometry but half the mass, the low
mass bar. However in these cases our simulations have lower resolution
than the default bar case (see Sect.~\ref{sect:sim}). In order to get
the same number of objects as the \RAVE red clump sample, we mirror
the particles above and below the $z = 0$ plane (so that we double the
resolution). However, we only do this for those
$\left(\alpha,\delta,K\right)$ bins that contain fewer particles than
the observed number of red clump stars in \emph{RAVE}. We are allowed
to do this because the potential is symmetric with the respect of the
Galactic plane and the same is true for our test particle simulations
(at least when they reach a steady state). After this operation, the
discrepancy in total number of objects between \RAVE and the
simulation is smaller than $1.5\%$. The regions slightly
underpopulated are those with $0\lesssim\alpha\lesssim 80\degree$,
$320\degree\lesssim\alpha\lesssim360\degree$ and
$\delta\lesssim-60\degree$. In what follows we only consider the
standard Sun's location, namely the case with the Sun at
$(R,\phi,z)=(8\Kpc,-20\degree,0)$.

\subsubsection{$R=8\Kpc$, $\phi=-20\degree$, long bar}

The long bar has a stronger effect than the default bar near the Sun,
because its gravitational force is larger in the solar neighborhood
(see M13). This is evident looking at the black line in
Fig.~\ref{fig:trends_LB_8kpc_20deg}, but also in the red and blue
curves which are obtained after applying the \RAVE selection function
and error convolution. Moreover, for $|z|>0.5\Kpc$, $\avvR$ grows
steeply for $R\lesssim 7.5-8\Kpc$ and decreases steeply for
$R\gtrsim7.5-8\Kpc$. In the central slices the effect of the selection
function and error enhances the gradient, for the same reasons as in
the standard case. For $|z|>0.5\Kpc$ the selection function together
with the errors preferentially pick out bins with smaller $R$, where
$\avvR$ increases, which results in washing out the gradient.

\subsubsection{$R=8\Kpc$, $\phi=-20\degree$, less massive bar}

As shown from the black lines in Fig.~\ref{fig:trends_LMB_8kpc_20deg}
and not surprisingly, the gradient induced by the less massive bar is
shallower than the default bar because the bar is weaker. Formally the
force of less massive bar is half that of the default bar. However the
non-axisymmetric part of the force (i.e., excluding the monopole term
associated to the bar) only differs by $\sim30\%$ in the solar
neighborhood (see M13).

Once the selection function and the error convolution have been
applied, the resulting $v_R$ gradients are significantly shallower
almost everywhere.

\section{Reasons for the velocity gradient}\label{sect:expl}
\begin{figure}
  \centering
  \includegraphics[height=0.8\textheight]{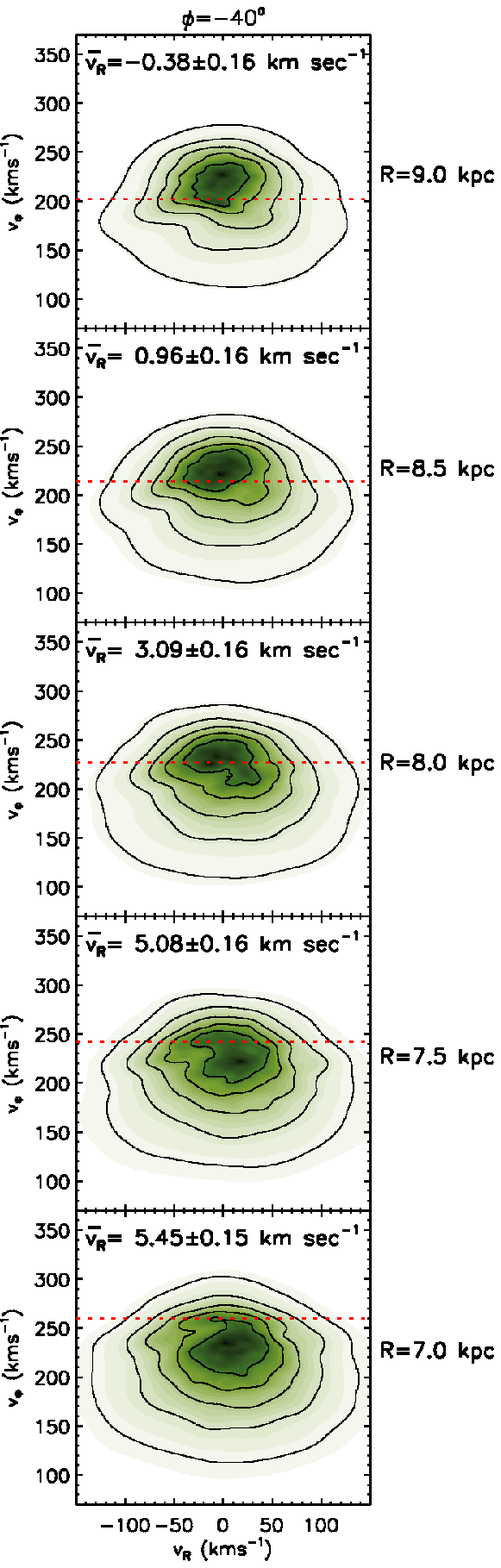}
  \caption{Velocity distribution in cylinders of radius $300\pc$ and
    height $600\pc$, centered at $\phi=-40\degree$, $z=0$ and
    different radii for the default bar case, inside the $3\Kpc$
    sphere centered at $(R,\phi,z)=(8\Kpc,-40\degr,0)$. The density
    distribution is obtained with an adaptive kernel estimator (see
    M13).}
  \label{fig:vRvphi}
\end{figure}
To explain the way the bar can induce a large scale radial velocity
gradient in the Milky Way, as in the simulations discussed here, we
show Fig.~\ref{fig:vRvphi}. In this figure we have plotted the
velocity distribution in the $v_R$ and $v_\phi$ components of the
simulation with the default bar at $\phi=-40\degree$ and different
$R$, inside small cylinders of radius $300\pc$ and height $600\pc$,
centered on the Galactic plane. The density field is estimated with an
adaptive kernel estimator (see details in M13).

Let us consider an axisymmetric potential similar to the one of a disk
galaxy. 
A star with angular momentum $L_z=Rv_\phi$ is associated with a
circular orbit with radius $\Rg$, the ``guiding center'', such that
$L_z=\Rg^2\Omega\left(\Rg\right)$. Therefore, when it passes from $R$,
its tangential velocity is
\begin{equation}
  v_\phi=\frac{\Rg^2\Omega\left(\Rg\right)}{R}.
\end{equation}
Near the Sun, the bar most strongly influences the stars with
$\Rg=\ROLR$. This is shown in Fig.~\ref{fig:vRvphi}, where the red
dashed line denotes
$\vOLR$, which is $v_\phi$ of orbits that have $\Rg=\ROLR$, computed
using the monopole component of the Fourier decomposition of the
potential in $\phi$ and where $R$ is taken at the center of the
volumes.  In fact, we note that around
$\vOLR$ the velocity distribution is split in two parts: the particles
with $v_\phi>\vOLR$ have $\avvR \lesssim 0$, the particles with
$v_\phi<\vOLR$ have $\avvR>0$ (\citealt{Kalnajs1991} introduced the
idea that the outer Lindblad resonance could account for bifurcation
of the solar neighborhood velocity distribution). We dub the former
group ``LSR mode'' and the latter ``OLR mode'', in the same fashion of
\cite{Dehnen2000}, that linked the latter to the Hercules stream. The
division is particularly clear for the volume centered at $R=8\Kpc$.

A first order treatment of nearly circular orbits in a weak bar
potential (\citealt{BT2008}, Sect. 3.3.3) shows that the bar
gravitational force stretches these orbits in two directions near the
outer Lindblad resonance and in the frame of reference of the bar: the
orbits with $\Rg<\ROLR$ are stretched perpendicular to the long axis
of the bar and the orbits with $\Rg>\ROLR$ are aligned parallel to the
long axis of the bar. The nearer $\Rg$ to $\ROLR$, the stronger the
effect. When they pass near the Sun, the orbits with $\Rg<\ROLR$
($\Rg>\ROLR$) have positive (negative) $v_R$\footnote{This prediction
  of the first order treatment can be obtained from the time
  derivative of Eq.~(3.148a) of \cite{BT2008}, and is confirmed in our
  simulations.}. Orbits with $\Rg$ far enough from $\ROLR$ are not
very affected by the bar, and on average have $v_R\sim0$. We see this
reflected in Fig.~\ref{fig:vRvphi}: the OLR mode is formed by stars
with $\Rg<\ROLR$, and the LSR mode by stars with $\Rg>\ROLR$.

When the volume is centered near $\ROLR$ (e.g., $R=7.5-8\Kpc$ in
Fig.~\ref{fig:vRvphi}), the orbits of the OLR mode with $\Rg<\ROLR$
dominate the velocity distribution of the particles, resulting in
$\avvR>0$ for the whole volume. As we go further from the Outer
Lindblad Resonance less particles populate the OLR mode. In
particular, if we only consider volumes centered at $R>\ROLR$, this
results in a negative $\avvR$ gradient (positive for $R<\ROLR$). This
is why we observe $\avvR$ gradients and a double behavior inside and
outside the outer Lindblad resonance.

Note that, because of the symmetry of the problem, for volumes
centered at positive $\phi$ the situation is reversed: $\avvR$ of the
OLR mode is negative and the gradient is positive. 

\section{Discussion}\label{sect:disc}
\begin{figure*}
  \centering
  \includegraphics[width=\textwidth]{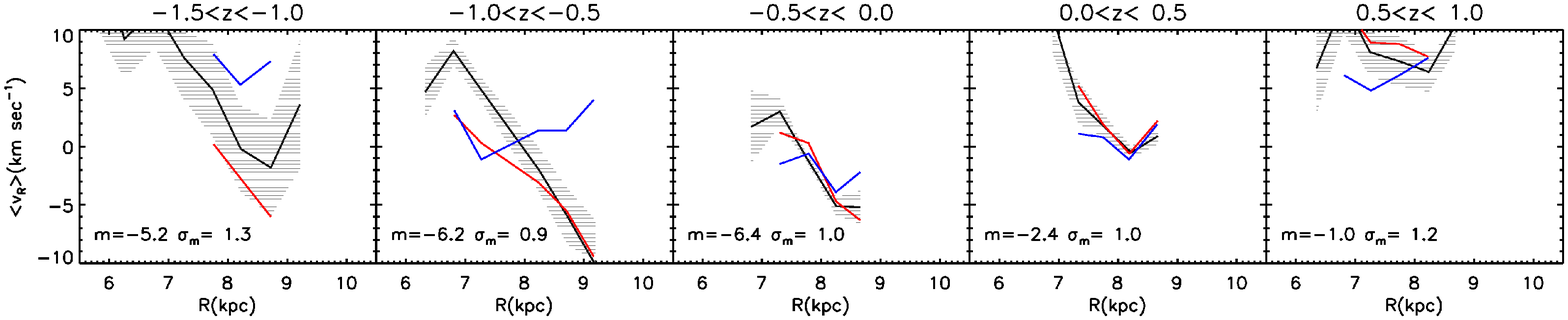}
  \caption{As in Fig.~\ref{fig:trends_GB_8kpc_20deg}, but for the
    \RAVE red clump stars used in W13. The shaded areas represent the
    measurement errors, the blue line the results obtained with the
    UCAC3 proper motions, the red line the results obtained with the
    SPM4 proper motions (see W13).}
  \label{fig:Will}
\end{figure*}

In this Section we compare our results in the case where the selection
function and the error convolution are applied (red and blue lines in
Figs.~\ref{fig:trends_GB_8kpc_20deg} -
\ref{fig:trends_LMB_8kpc_20deg}, that are similar) with the
findings of W13 and in particular with the $v_R$ trends in their
Fig.~8 (here reproduced in Fig.~\ref{fig:Will}). Depending on the
assumed Sun's motion with the respect of the Local Standard of Rest,
the curves may shift up or down in $v_R$, but the overall trends
remain unaffected (as shown in Fig.~9 in W13). Note that we have not
included a correction for the solar motion in our analysis so
far. Nonetheless, the $v_R$ values are comparable to those of W13.

However, only a qualitative comparison is warranted, as our simulation
does not really reproduce in detail the properties of the Milky Way
(e.g, the rotation curve is falling off near the Sun, the peak
velocity is larger than observed, etc). Furthermore, as we have noted,
after the error convolution and \RAVE selection function are applied,
the underlying trends are sometimes modified, implying that care
should be taken to avoid over interpretation of the results.

An important difference is that the radial velocity gradients found in
\RAVE (Fig.~\ref{fig:Will}) are much steeper than in any of our models
(Fig.~\ref{fig:trends_GB_8kpc_20deg} -
\ref{fig:trends_LMB_8kpc_20deg}), except perhaps for the slices with
$z > 0$. In magnitude, the model gradients resemble more the low limit
of the S11 estimate, i.e., $\frac{\de\avvR}{\de
  R}\gtrsim3\kmseckpc$. The trends are also different in the sense
that most of the cases we have explored show a flat/increasing part
(e.g., at $R < 7.5\Kpc$ for the default case) followed by a steeper
decline at larger radii, a behavior that seems to be absent (or is not
as clear) in the data as shown in Fig.~\ref{fig:Will}.

Although as stated above, in absolute terms the actual values of $v_R$
depend on the solar motion, in Fig.~10 we note that the mean value of
$v_R$ changes with distance from the plane, when averaged over the
whole radial distance range. This behavior is also present, and in the
same sense, in our default model, where for the three $z$ slices at $z
> −0.5 \Kpc$: in the central bins $\avvR > 0$ for $R < 8 \Kpc$ and
$v_R \gtrsim 0$ for $R > 8 \Kpc$, and $\avvR > 0$ everywhere for $z >
0.5 \Kpc$.

\begin{figure*}
  \centering
  \includegraphics[width=\textwidth]{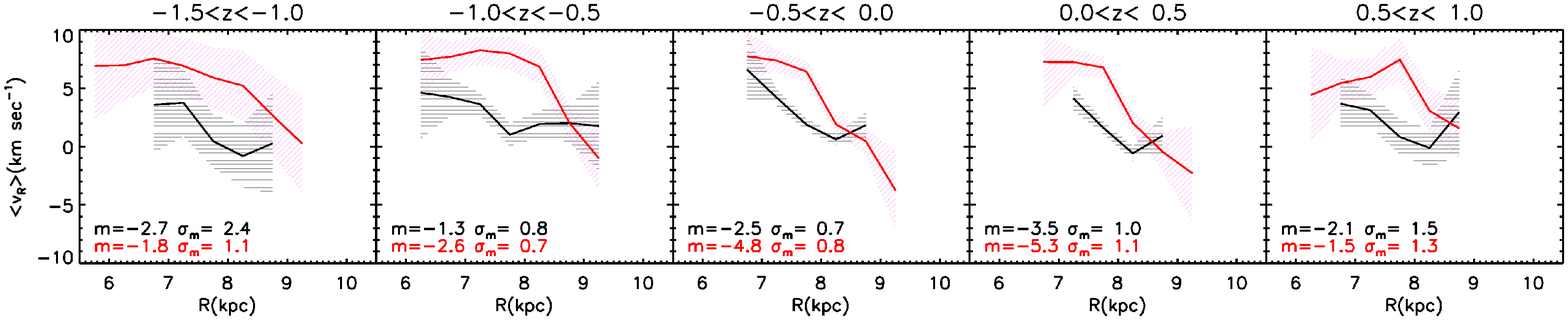}
  \includegraphics[width=\textwidth]{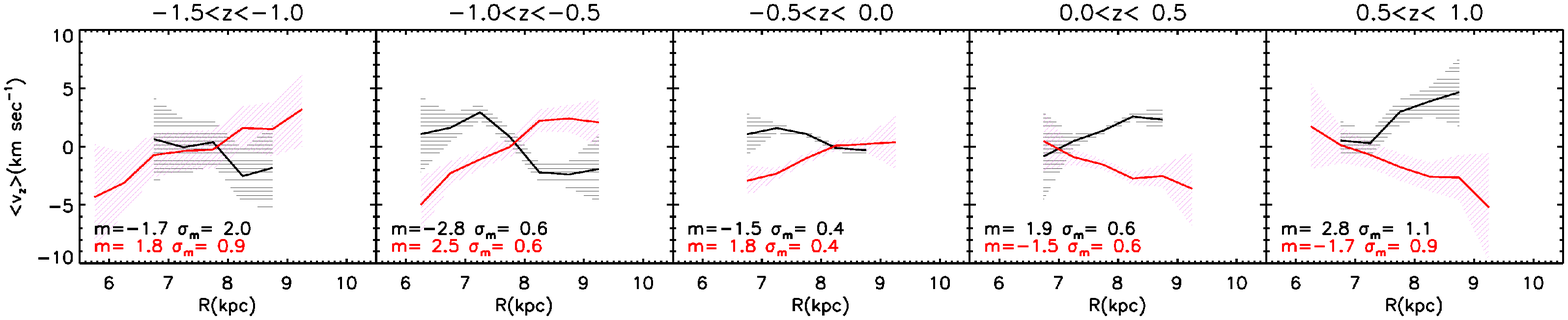}
  \caption{Effect of a systematic errors in the distance determination
    on the $\avvR$ (top row) and $\avvz$ (bottom row) trends in the
    default bar case. The simulation is also convolved with the random
    errors. Black lines: $20\%$ systematic underestimation of the
    distances. Red lines: $20\%$ systematic overestimation of the
    distances.}
  \label{fig:trends_syst}
\end{figure*}

Systematic errors in the distances (and more specifically the assumed
absolute magnitude for the red clump stars) could also affect our
results. However, we find that when we use the other red clump
magnitude normalizations considered in W13 ($\MK = −1.54$ and
$\MK=−1.64 + 0.0625|z(\mathrm{kpc})|$) our conclusions are not
affected, because these only induce small distance changes ($\sim
5\%$). On the other hand, if the distances were more significantly
overestimated, by for example $20\%$, the $v_R$ gradients would become
steeper as shown by the red curves in the top panels of
Fig.~\ref{fig:trends_syst}, while an underestimation by the same
magnitude would lead to shallower gradients as shown by the black
curves in the figure. Interestingly, in the case that the distances
are overestimated a gradient in $v_z$ with radius is also induced as
shown in the bottom panels of Fig.~\ref{fig:trends_syst}, which is
positive below the plane, and negative above the plane, i.e. in the
same sense as found by W13. Since in the literature
(\citealt{Alves2000,GrocholskiSarajedini2002,SalarisGirardi2002,Groenewegen2008})
most other estimates of the red clump magnitude are fainter than what
we have assumed, it may seem more likely that the distances in W13
have been systematically overestimated than underestimated\footnote{In
  fact, \cite{Binney2014} recently estimated the absolute magnitude of
  the \RAVE red clump stars to be $\sim 0.1~\mathrm{mag}$ fainter than
  in W13.}, however not as extremely as we have tested in these last
examples.

It should be noted that the fact that the models explored in this work
do not fit the observed data, does not necessarily imply that the
Galactic bar is not the cause for the observed gradients. With our
models we only have explored a very small portion of a large parameter
space and, for example, steeper gradients can be obtained by
increasing the bar strength near the Sun, or even with a circular
speed curve falling more slowly with $R$, at fixed bar pattern
speed\footnote{This can be shown in 2D with a simple potential with
  power-law velocity curve, applying \cite{Kuijken1991} theory for the
  behavior of the mean velocities under the influence of a
  non-axisymmetric perturbation of multipole order $m$ (see also
  \citealt{Muhlbauer2003}).}. Since some of the kinematic substructure
in the solar neighborhood can be explained by the bar, we expect that
its dynamical effect (for the current bar parameters estimates) should
at least be partly responsible for the observed negative $\avvR$
gradients.

\section{Conclusions}\label{sect:concl}
In this work we have proposed a new explanation for the recent
discovery (S11, W13) of a negative $R$ gradient of the
(galactocentric) radial velocity. We found that the bar can create a
negative gradient if the Sun is placed just outside the outer Lindblad
resonance and at angles from the long axis of the bar similar to the
current estimates from the literature. The velocity gradients become
steeper when increasing the angle from the bar and also for the Long
Bar model. On the other hand, in the less massive bar case they become
shallower. Moreover, such gradients do not depend strongly on the
height from the Galactic plane. This happens because the bar affects
the kinematics of the Galaxy almost in the same way from $z=0$ to
$z\sim2\Kpc$, as explained in M13. Because of this, the bar provides a
natural mechanism for the observed gradients at different heights.

We compared the 3D test particle simulations presented in M13 with the
findings of \emph{RAVE}, after applying the \emph{RAVE} selection
function and proper error convolution. The gradients exist in our
simulations for all bar parameters and positions of the Sun explored
(all outside the outer Lindblad resonance). These gradients are never
completely washed out by the selection function and the errors, but
rather they are enhanced in some cases. In fact, the gradients in the
solar neighborhood spheres considered are in general shallower than
those observed in the Milky Way, but the selection function can
enhance them to the level of $\sim3-4\kmseckpc$ (as e.g., happens for
all the studied simulation slices with $|z|<0.5\Kpc$ and the Sun
centered at $R=8\Kpc$).

However, none of the models that we explored in this work accurately
describes the behavior in \RAVE of $\avvR$ at every $z$: the gradients
are too shallow for $z<0$. Some models resemble \RAVE for $z>0$,
especially our default bar case at $R=8\Kpc$ and
$\phi\leq-20\degr$. We conclude from this that the bar should at least
contribute to the negative gradient observed, for position angles with
respect to the bar $\phi < 0$ and for locations of the Sun near but
outside the outer Lindblad resonance ($R > \ROLR$).

Furthermore, our simulations do not show any kind of vertical velocity
gradient as seen in the data for \RAVE by W13. This result is
consistent with the distribution function of the simulated disks being
an even function of $v_z$. On the other hand, the recent paper by
\cite{Faure2014} shows that a 3D model for spiral arms is successful
in reproducing radial and vertical velocity gradients similar to those
observed in W13. In reality both effects of bar and spiral arms
probably coexist and shape the velocity distribution of the solar
neighborhood. However, while in the case of the bar the slope of the
radial velocity gradient depends significantly on the angular location
of the observer in the Galaxy, in the case of tightly wound spirals
the angle is much less important (Fig.~6 and 7 in
\citealt{Faure2014}). Future observations of the Galactic disk (e.g.,
obtained with the \emph{Gaia} satellite) are expected to be
sufficiently extended to distinguish whether the main cause of the
radial velocity gradient is the bar or the spiral arms.

A natural future development of this work is to fit the kinematics of
the extended solar neighborhood with the analytic predictions from the
bar perturbation theory, in the same fashion as in \cite{Siebert2012}
for the spiral arms, in order to retrieve the best fit values for the
bar pattern speed, bar angle, and bar strength.

\begin{acknowledgements}
  We would like to thank T. Piffl and M. Williams for useful
  discussions. We also thank M. Williams for kindly providing the
  \emph{RAVE} data used in \cite{Williams2013}. The authors gratefully
  acknowledge support from the European Research Council under ERC
  Starting Grant GALACTICA-240271.
\end{acknowledgements}

\bibliography{Monaribib}{}

\begin{thebibliography}{30}
\expandafter\ifx\csname natexlab\endcsname\relax\def\natexlab#1{#1}\fi

\bibitem[{{Alves}(2000)}]{Alves2000}
{Alves}, D.~R. 2000, \apj, 539, 732

\bibitem[{{Antoja} {et~al.}(2008){Antoja}, {Figueras}, {Fern{\'a}ndez}, \&
  {Torra}}]{Antoja2008}
{Antoja}, T., {Figueras}, F., {Fern{\'a}ndez}, D., \& {Torra}, J. 2008, \aap,
  490, 135

\bibitem[{{Antoja} {et~al.}(2011){Antoja}, {Figueras}, {Romero-G{\'o}mez},
  {Pichardo}, {Valenzuela}, \& {Moreno}}]{Antoja2011}
{Antoja}, T., {Figueras}, F., {Romero-G{\'o}mez}, M., {et~al.} 2011, \mnras,
  418, 1423

\bibitem[{{Antoja} {et~al.}(2012){Antoja}, {Helmi}, {Bienayme},
  {Bland-Hawthorn}, {Famaey}, {Freeman}, {Gibson}, {Gilmore}, {Grebel},
  {Minchev}, {Munari}, {Navarro}, {Parker}, {Reid}, {Seabroke}, {Siebert},
  {Siviero}, {Steinmetz}, {Williams}, {Wyse}, \& {Zwitter}}]{Antoja2012}
{Antoja}, T., {Helmi}, A., {Bienayme}, O., {et~al.} 2012, \mnras, 426, L1

\bibitem[{{Antoja} {et~al.}(2009){Antoja}, {Valenzuela}, {Pichardo}, {Moreno},
  {Figueras}, \& {Fern{\'a}ndez}}]{Antoja2009}
{Antoja}, T., {Valenzuela}, O., {Pichardo}, B., {et~al.} 2009, \apjl, 700, L78

\bibitem[{{Bessell} \& {Brett}(1989)}]{BesselBrett1989}
{Bessell}, M.~S. \& {Brett}, J.~M. 1989, in Lecture Notes in Physics, Berlin
  Springer Verlag, Vol. 341, Infrared Extinction and Standardization, ed. E.~F.
  {Milone}, 61

\bibitem[{{Binney} {et~al.}(2014){Binney}, {Burnett}, {Kordopatis}, {McMillan},
  {Sharma}, {Zwitter}, {Bienaym{\'e}}, {Bland-Hawthorn}, {Steinmetz},
  {Gilmore}, {Williams}, {Navarro}, {Grebel}, {Helmi}, {Parker}, {Reid},
  {Seabroke}, {Watson}, \& {Wyse}}]{Binney2014}
{Binney}, J., {Burnett}, B., {Kordopatis}, G., {et~al.} 2014, \mnras, 437, 351

\bibitem[{{Binney} \& {Tremaine}(2008)}]{BT2008}
{Binney}, J. \& {Tremaine}, S. 2008, {Galactic Dynamics: Second Edition}, ed.
  {Binney, J.~\& Tremaine, S.} (Princeton University Press)

\bibitem[{{Bissantz} \& {Gerhard}(2002)}]{BissantzGerhard2002}
{Bissantz}, N. \& {Gerhard}, O. 2002, \mnras, 330, 591

\bibitem[{{De Simone} {et~al.}(2004){De Simone}, {Wu}, \&
  {Tremaine}}]{DeSimone2004}
{De Simone}, R., {Wu}, X., \& {Tremaine}, S. 2004, \mnras, 350, 627

\bibitem[{{Dehnen}(1998)}]{Dehnen1998}
{Dehnen}, W. 1998, \aj, 115, 2384

\bibitem[{{Dehnen}(2000)}]{Dehnen2000}
{Dehnen}, W. 2000, \aj, 119, 800

\bibitem[{{Famaey} {et~al.}(2005){Famaey}, {Jorissen}, {Luri}, {Mayor}, {Udry},
  {Dejonghe}, \& {Turon}}]{Famaey2005}
{Famaey}, B., {Jorissen}, A., {Luri}, X., {et~al.} 2005, \aap, 430, 165

\bibitem[{{Faure} {et~al.}(2014){Faure}, {Siebert}, \& {Famaey}}]{Faure2014}
{Faure}, C., {Siebert}, A., \& {Famaey}, B. 2014, \mnras, 440, 2564

\bibitem[{{Ferrers}(1870)}]{Ferrers1870}
{Ferrers}, N.~M. 1870, Royal Society of London Philosophical Transactions
  Series I, 160, 1

\bibitem[{{Fux}(2001)}]{Fux2001}
{Fux}, R. 2001, \aap, 373, 511

\bibitem[{{Grocholski} \& {Sarajedini}(2002)}]{GrocholskiSarajedini2002}
{Grocholski}, A.~J. \& {Sarajedini}, A. 2002, \aj, 123, 1603

\bibitem[{{Groenewegen}(2008)}]{Groenewegen2008}
{Groenewegen}, M.~A.~T. 2008, \aap, 488, 935

\bibitem[{{Kalnajs}(1991)}]{Kalnajs1991}
{Kalnajs}, A.~J. 1991, in Dynamics of Disc Galaxies, ed. {B.~Sundelius}, 323--+

\bibitem[{{Kuijken} \& {Tremaine}(1991)}]{Kuijken1991}
{Kuijken}, K. \& {Tremaine}, S. 1991, in Dynamics of Disc Galaxies, ed.
  B.~{Sundelius}, 71

\bibitem[{{Mayor}(1970)}]{Mayor1970}
{Mayor}, M. 1970, \aap, 6, 60

\bibitem[{{Monari} {et~al.}(2013){Monari}, {Antoja}, \& {Helmi}}]{Monari2013}
{Monari}, G., {Antoja}, T., \& {Helmi}, A. 2013, arXiv:1306.2632

\bibitem[{{M{\"u}hlbauer} \& {Dehnen}(2003)}]{Muhlbauer2003}
{M{\"u}hlbauer}, G. \& {Dehnen}, W. 2003, \aap, 401, 975

\bibitem[{{Quillen} {et~al.}(2011){Quillen}, {Dougherty}, {Bagley}, {Minchev},
  \& {Comparetta}}]{Quillen2011}
{Quillen}, A.~C., {Dougherty}, J., {Bagley}, M.~B., {Minchev}, I., \&
  {Comparetta}, J. 2011, \mnras, 417, 762

\bibitem[{{Salaris} \& {Girardi}(2002)}]{SalarisGirardi2002}
{Salaris}, M. \& {Girardi}, L. 2002, \mnras, 337, 332

\bibitem[{{Siebert} {et~al.}(2012){Siebert}, {Famaey}, {Binney}, {Burnett},
  {Faure}, {Minchev}, {Williams}, {Bienaym{\'e}}, {Bland-Hawthorn}, {Boeche},
  {Gibson}, {Grebel}, {Helmi}, {Just}, {Munari}, {Navarro}, {Parker}, {Reid},
  {Seabroke}, {Siviero}, {Steinmetz}, \& {Zwitter}}]{Siebert2012}
{Siebert}, A., {Famaey}, B., {Binney}, J., {et~al.} 2012, \mnras, 425, 2335

\bibitem[{{Siebert} {et~al.}(2011{\natexlab{a}}){Siebert}, {Famaey}, {Minchev},
  {Seabroke}, {Binney}, {Burnett}, {Freeman}, {Williams}, {Bienaym{\'e}},
  {Bland-Hawthorn}, {Campbell}, {Fulbright}, {Gibson}, {Gilmore}, {Grebel},
  {Helmi}, {Munari}, {Navarro}, {Parker}, {Reid}, {Siviero}, {Steinmetz},
  {Watson}, {Wyse}, \& {Zwitter}}]{Siebert2011grad}
{Siebert}, A., {Famaey}, B., {Minchev}, I., {et~al.} 2011{\natexlab{a}},
  \mnras, 412, 2026

\bibitem[{{Siebert} {et~al.}(2011{\natexlab{b}}){Siebert}, {Williams},
  {Siviero}, {Reid}, {Boeche}, {Steinmetz}, {Fulbright}, {Munari}, {Zwitter},
  {Watson}, {Wyse}, {de Jong}, {Enke}, {Anguiano}, {Burton}, {Cass}, {Fiegert},
  {Hartley}, {Ritter}, {Russel}, {Stupar}, {Bienaym{\'e}}, {Freeman},
  {Gilmore}, {Grebel}, {Helmi}, {Navarro}, {Binney}, {Bland-Hawthorn},
  {Campbell}, {Famaey}, {Gerhard}, {Gibson}, {Matijevi{\v c}}, {Parker},
  {Seabroke}, {Sharma}, {Smith}, \& {Wylie-de Boer}}]{Siebert2011b}
{Siebert}, A., {Williams}, M.~E.~K., {Siviero}, A., {et~al.}
  2011{\natexlab{b}}, \aj, 141, 187

\bibitem[{{Steinmetz} {et~al.}(2006){Steinmetz}, {Zwitter}, {Siebert},
  {Watson}, {Freeman}, {Munari}, {Campbell}, {Williams}, {Seabroke}, {Wyse},
  {Parker}, {Bienaym{\'e}}, {Roeser}, {Gibson}, {Gilmore}, {Grebel}, {Helmi},
  {Navarro}, {Burton}, {Cass}, {Dawe}, {Fiegert}, {Hartley}, {Russell},
  {Saunders}, {Enke}, {Bailin}, {Binney}, {Bland-Hawthorn}, {Boeche}, {Dehnen},
  {Eisenstein}, {Evans}, {Fiorucci}, {Fulbright}, {Gerhard}, {Jauregi}, {Kelz},
  {Mijovi{\'c}}, {Minchev}, {Parmentier}, {Pe{\~n}arrubia}, {Quillen}, {Read},
  {Ruchti}, {Scholz}, {Siviero}, {Smith}, {Sordo}, {Veltz}, {Vidrih}, {von
  Berlepsch}, {Boyle}, \& {Schilbach}}]{Steinmetz2006}
{Steinmetz}, M., {Zwitter}, T., {Siebert}, A., {et~al.} 2006, \aj, 132, 1645

\bibitem[{{Williams} {et~al.}(2013){Williams}, {Steinmetz}, {Binney},
  {Siebert}, {Enke}, {Famaey}, {Minchev}, {de Jong}, {Boeche}, {Freeman},
  {Bienaym{\'e}}, {Bland-Hawthorn}, {Gibson}, {Gilmore}, {Grebel}, {Helmi},
  {Kordopatis}, {Munari}, {Navarro}, {Parker}, {Reid}, {Seabroke}, {Sharma},
  {Siviero}, {Watson}, {Wyse}, \& {Zwitter}}]{Williams2013}
{Williams}, M.~E.~K., {Steinmetz}, M., {Binney}, J., {et~al.} 2013, \mnras,
  436, 101

\end{thebibliography}
\bibliographystyle{aa}

\end{document}